\documentclass[conference]{IEEEtran}
\IEEEoverridecommandlockouts

\usepackage{cite}
\usepackage{amsmath,amssymb,amsfonts}
\usepackage{algorithmic}
\usepackage{graphicx}
\usepackage{textcomp}
\usepackage{xcolor}

\usepackage{tabularx}
\usepackage{booktabs}
\usepackage[draft]{hyperref}
\usepackage{array}

\usepackage{eso-pic}
\newcommand{\ieeecopyrightyear}{2026}
\newcommand{\ieeecopyrightnotice}{%
  \AddToShipoutPictureBG*{%
    \AtPageLowerLeft{%
      \put(0,0.6cm){\makebox[\paperwidth]{%
        \parbox{0.86\paperwidth}{\footnotesize\centering
        \copyright~\ieeecopyrightyear{} IEEE. Personal use of this material is permitted. Permission from IEEE must be obtained for all other uses, in any current or future media, including reprinting/republishing this material for advertising or promotional purposes, creating new collective works, for resale or redistribution to servers or lists, or reuse of any copyrighted component of this work in other works.}}}}}}

\definecolor{cqwen25}{HTML}{4C72B0}
\definecolor{cvoxtral}{HTML}{DD8452}
\definecolor{cqwen3o}{HTML}{8172B3}
\definecolor{cgood}{HTML}{2F8F4E}
\newcommand{\swone}[1]{{\color{#1}\rule[-0.2ex]{1.7mm}{1.7mm}}}
\newcommand{\swtwo}[2]{{\color{#1}\rule[-0.2ex]{0.85mm}{1.7mm}}{\color{#2}\rule[-0.2ex]{0.85mm}{1.7mm}}}
\newcommand{\cmark}{\textcolor{cgood}{\checkmark}}
\newcommand{\dna}{\textcolor{gray}{--}}

\makeatletter
\newcommand{\linebreakand}{%
  \end{@IEEEauthorhalign}
  \hfill\mbox{}\par
  \mbox{}\hfill
  \begin{@IEEEauthorhalign}
}
\makeatother

\def\BibTeX{{\rm B\kern-.05em{\sc i\kern-.025em b}\kern-.08em
    T\kern-.1667em\lower.7ex\hbox{E}\kern-.125emX}}
\begin{document}

\title{The \textbf{Be}yond \textbf{Tra}nscription \textbf{C}hallenge: \\End-to-End Long Clinical Audio Summarization\\
}

\author{\IEEEauthorblockN{Jiyun Chun}
\IEEEauthorblockA{\textit{Computer Science and Engineering} \\
\textit{The Ohio State University}\\
Columbus, United States \\
chun.203@osu.edu}
\and
\IEEEauthorblockN{Samuele Cornell}
\IEEEauthorblockA{\textit{Language Technologies Institute} \\
\textit{Carnegie Mellon University}\\
Pittsburgh, USA \\
samuele.cornell@ieee.org}
\and
\IEEEauthorblockN{Syed-Amad Hussain}
\IEEEauthorblockA{\textit{Abigail Wexner Research Institute} \\
\textit{Nationwide Children's Hospital}\\
Columbus, Ohio \\
https://orcid.org/0000-0002-7529-8644}
\linebreakand
\IEEEauthorblockN{Leibny Paola Garcia Perera}
\IEEEauthorblockA{\textit{Center for Language and Speech Processing} \\
\textit{Johns Hopkins University}\\
Baltimore, United States \\
https://orcid.org/0000-0002-7449-5726}
\and
\IEEEauthorblockN{Ahmed Hassoon}
\IEEEauthorblockA{\textit{Data Science and AI Institute} \\
\textit{Johns Hopkins University}\\
Baltimore, United States \\
https://orcid.org/0000-0002-4930-3331}
\and
\IEEEauthorblockN{Markus Müller}
\IEEEauthorblockA{\textit{IAR} \\
\textit{Karlsruhe Institute of Technology}\\
Karlsruhe, Germany \\
https://orcid.org/0009-0000-1701-0019}
\linebreakand
\IEEEauthorblockN{Thomas Schaaf}
\IEEEauthorblockA{\textit{Health Information Systems} \\
\textit{Solventum}\\
Pittsburgh, United States \\
https://orcid.org/0000-0002-9569-4759}
\and
\IEEEauthorblockN{Andrew Perrault}
\IEEEauthorblockA{\textit{Computer Science and Engineering} \\
\textit{The Ohio State University}\\
Columbus, United States \\
perrault.17@osu.edu}
}

\ieeecopyrightnotice
\maketitle

\begin{abstract}
Recent open large audio language models perform well on short audio but struggle with long, multi-speaker conversations.
Clinical documentation is a useful setting for evaluating this ability---in doctor–patient conversations, symptoms, medical history, examination findings, clinical assessments, and treatment plans are distributed across the conversation.
In this setting, end-to-end models that generate SOAP notes directly from audio often add unsupported information or omit relevant details, and cascaded ASR-then-summarization pipelines are commonly reported to outperform them.
We introduce the \textbf{Beyond Transcription Challenge (BeTraC)} to address this gap. Participating systems must generate SOAP notes directly from long doctor–patient conversation audio without intermediate transcription.
BeTraC includes a Lightweight Track for a single end-to-end model with at most 6 billion parameters and a Heavyweight Track that allows multi-model systems, tools, and agent-based architectures with at most 36 billion parameters in total. 
Training data, in the form of synthetic doctor-patient conversation audio and SOAP pairs (Synth-DoPaCo), was provided to the teams.
Official rankings used Open Medical Concept F1 on held-out Synth-DoPaCo,
while post-competition evaluation assessed Faithfulness, Coverage,
Structure, and Conciseness with an LLM judge and tested generalization
to recorded OSCE interviews. Results showed strong transfer from synthetic to real data under both metrics. The two metrics agreed closely for high-scoring systems but diverged for low-scoring ones, exposing systematic blind spots in the automated metric.

\end{abstract}


\section{\textbf{Introduction}}
Recent open large audio language models (LALMs)~\cite{Qwen2.5-Omni,goel2025audioflamingo3advancing,Intelligence2025,li2025baichuanomni15technicalreport,xu2025qwen3,liu2025voxtral,arora2025landscape} have advanced rapidly in speech recognition, audio question answering, and voice based interaction~\cite{ghosh2025audio,ahia2025blabbrutallylongaudio,chen2025audiollmsreallylisten,kumar2025mmauprochallengingcomprehensivebenchmark}, suggesting that they may be able to understand long spoken interactions in real world settings. However, existing long audio benchmarks show that performance declines sharply as input duration increases~\cite{kumar2025mmauprochallengingcomprehensivebenchmark,ahia2025blabbrutallylongaudio}. Clinical documentation is an important application of this technology. In doctor patient conversations, symptoms, medical history, examination findings, clinical assessments, and treatment plans are often distributed across several minutes of conversation. Generating a useful \textit{\textbf{S}ubjective, \textbf{O}bjective, \textbf{A}ssessment, \textbf{P}lan} (SOAP) note therefore requires speech recognition, long context understanding, clinical information extraction, and structured generation. Compared with other long audio summarization tasks such as meeting summarization~\cite{carletta2005ami,rennard2023abstractive,cornell2025recent}, clinical documentation is more structured and therefore easier to evaluate. This structure has also motivated related shared tasks. MEDIQA Chat 2023~\cite{ben-abacha-etal-2023-overview} addressed clinical note generation from text based doctor patient conversations and synthetic conversation generation from clinical notes.

Clinical audio documentation typically relies on an ASR to text cascade. An ASR model transcribes the doctor patient conversation, and a separate LLM converts the transcript into a SOAP note. Although cascades generally outperform models that generate SOAP notes directly from audio, both approaches remain limited for the small open weight models considered in this challenge. Cascades also make it difficult to separate audio understanding from text summarization. In addition, ASR errors propagate to the final note, while transcription removes clinically meaningful nonlexical cues such as prosody, hesitation, stress, coughing, and breathing difficulty. Faithfully using these cues remains a longer term goal because current audio synthesis cannot fully reproduce them. BeTraC therefore focuses on narrowing the gap in generating clinically grounded SOAP notes with end to end models.

This gap leads to the main research question of the \textbf{Beyond Transcription Challenge (BeTraC)}\footnote{\url{https://betrac.github.io/}}: can open-weight LALM systems be constructed that generate faithful and clinically useful SOAP notes from doctor-patient conversation audio without using intermediate transcription? 
In particular, if many clinical claims generated by existing end-to-end audio-to-SOAP systems are not well supported by the actual conversation, can we reduce these hallucinations? 
BeTraC is designed as an end-to-end clinical audio reasoning challenge and enforces a no transcription rule. During inference, no textual representation of the audio, including a transcript, summary, or entity list, may pass between components. Only the final model may produce text derived from the audio.

Because privacy regulations such as HIPAA severely restrict the public release of real doctor patient conversation datasets\cite{s25144305, wang-etal-2024-notechat,papadopoulos-korfiatis-etal-2022-primock57,yim2023acibench,Fareez2022}, BeTraC uses Synth-DoPaCo~\cite{Labrak2026-yh}, an open synthetic dataset developed with clinicians for training and testing, together with automated evaluation. The dataset is generated entirely with open weight models. LLM role play produces the conversation transcript, the transcript is used to generate a SOAP note, and speech synthesis renders the conversation as audio. The official leaderboard is based on Open Medical Concept F1 on the Synth-DoPaCo test set. This metric combines MeSH keyword matching with named entity recognition using scispaCy~\cite{neumann-etal-2019-scispacy}.

These design choices raise questions about external validity because top systems may overfit the synthetic data or the Concept F1 metric. We examine these concerns through two independent post competition evaluations. We reevaluate the systems on 275 human acted doctor patient conversations, most of which are OSCE interviews~\cite{Fareez2022}, to assess overfitting to synthetic data, and evaluate every note with an LLM as a judge pipeline developed with clinicians~\cite{Labrak2026-yh} to assess overfitting to the metric.
The results show strong transfer from synthetic to real data, with system rankings on synthetic and real conversations closely aligned under both metrics. Concept F1 and the LLM judge also agree for high scoring systems but diverge for low scoring systems, revealing systematic blind spots in the automated metric.
Overall, BeTraC provides a standardized training set and benchmark for long form audio reasoning, faithful clinical documentation, hallucination analysis, and LALM evaluation. Our code will be released publicly.

\section{\textbf{Background}}

\subsection{\textbf{Synth-DoPaCo}}
BeTraC uses the Synth-DoPaCo dataset (Table~\ref{tab:dataset_statistics})~\cite{Labrak2026-yh} to evaluate the ability of LALMs to generate structured clinical documentation from long doctor--patient conversations. Synth-DoPaCo is a fully synthetic dataset of 8,800 first-visit encounters ($\sim$1,300 hours), each paired with an audio recording and a
reference SOAP note. The dataset was developed to address the lack of resources for long-form audio understanding and was generated entirely using open-weight models with permissive licenses.

\subsection{\textbf{Data Generation Pipeline}}
Synth-DoPaCo generates each example using a three stage pipeline built entirely with open weight models. It consists of persona driven dialogue generation using Gemma 3 27B IT~\cite{gemma3modelcard}, multi speaker audio synthesis with realistic acoustics and non speech events, and reference note generation from clinical facts extracted from the ground truth transcript using Kimi K2 Thinking~\cite{kimiteam2026kimik2openagentic}. Using only these facts helps limit unsupported clinical information. The pipeline is described fully in~\cite{Labrak2026-yh}. A 10 minute conversation contains only \textasciitilde1{,}200 text tokens but roughly ten times as many audio tokens, which contributes to the long context difficulty faced by end to end systems.

Three distinct open-weight models play roles in this paper: Gemma 3 27B IT generates the dialogues, Kimi K2 Thinking generates the reference SOAP notes, and, for the post-competition analysis (\S~\ref{sec:ood}), Gemma 4 31B IT~\cite{gemma4modelcard} serves as the LLM judge.

\begin{table}[t]
\centering
\caption{Statistics and release settings of the BeTraC datasets.}
\label{tab:dataset_statistics}

\scriptsize
\setlength{\tabcolsep}{2.5pt}
\renewcommand{\arraystretch}{0.95}

\begin{tabularx}{0.94\columnwidth}{@{}lrrX@{}}
\toprule
\textbf{Split} & \textbf{\#Conv.} & \textbf{Hrs} & \textbf{Release} \\
\midrule
Train~\cite{Labrak2026-yh}
& 7,200 & $\sim$1,087 & Audio + reference SOAP notes \\

Dev~\cite{Labrak2026-yh}
& 400 & $\sim$59 & Audio + reference SOAP notes \\

\textbf{Test}~\cite{Labrak2026-yh}
& 600 & $\sim$101 & Audio only\textsuperscript{*} \\

OSCE~\cite{Fareez2022}
& 272 & $\sim$55 & Audio only\textsuperscript{*} \\

Realistic
& 3 & $\sim$0.4 & Audio only\textsuperscript{*} \\
\bottomrule
\end{tabularx}

\vspace{1pt}
\begin{minipage}{0.94\columnwidth}
\scriptsize
\textsuperscript{*}Reference SOAP notes are withheld for blind evaluation.
\end{minipage}
\end{table}

\section{\textbf{The BeTraC Challenge}}
\subsection{\textbf{Task Input, Output, and Submission Format}}
Each BeTraC test instance consists of a doctor-patient conversation audio. Participating systems receive the raw audio as input and generate a SOAP note containing the clinical information from the conversation. Each audio file must be processed independently without using information from other test files.
The target output is a SOAP note with information organized into Subjective, Objective, Assessment, and Plan sections. The Subjective section contains patient reported symptoms and medical history. The Objective section contains observations and examination findings. The Assessment section contains the clinical assessment. The Plan section contains treatment and follow up plans.

Following the challenge timeline, teams were required to submit brief system descriptions by June 24, 2026 to receive the test audio. Teams then had approximately one week to generate and submit SOAP notes, with submissions due on July 1, 2026.
For each team that did so, the organizers provided them with 875 test audios in a random order, without reference notes. This consisted of 600 Synth-DoPaCo test set audios and 275 out-of-distribution (OOD) test audios, which are described in \S~\ref{sec:ood}. Teams were instructed not to modify their systems between receiving the test audio and submitting their generated SOAP notes. 
Each submission included the raw model output for every test audio file, a script that extracts the final SOAP note from the raw output, and the extracted SOAP notes. We required that the raw output contain every token generated by the model, including chain of thought, tags, delimiters, and any other generated content. Heavyweight systems that use multiple models or tools were required to submit a complete execution trace containing the inputs and outputs of all model and tool calls.

 \subsection{\textbf{Tracks and Eligibility}}

All submitted systems had to use approved open-weight models meeting four eligibility criteria: publicly available weights, a reproducible (pinned) checkpoint, a published model card or technical report, and a forward-pass parameter count within the track budget. Proprietary or API-only models were disallowed anywhere in the system, including for generating synthetic training data such as SOAP notes or TTS audio.

The defining constraint is the \emph{no-transcription rule}, which governs both what may pass between components at inference and how the final model may be trained.

\noindent\textbf{At inference.} No textual representation of the spoken content (e.g., transcript, paraphrase, summary, keyword or clinical-entity list, or draft note) may pass between separate models or components, and only the final model may emit text derived from the audio. This excludes conventional ASR cascades. ASR-pretrained encoders may still be used, but only as non-decoding acoustic feature extractors.

\noindent \textbf{At training.} Multi-task training and distillation are permitted, and the final model's chain-of-thought (CoT) is allowed and unconstrained in form, i.e., emergent transcript-like text is not a violation. The one restriction: the model must not have been trained with a transcript of the input audio as a CoT target, prefix, or supervision signal, nor distilled from a transcript-aligned teacher~\cite{Wei2022-af}.

\noindent \textbf{Prohibited.} Post-hoc learned rewriting of the note, cross-file context, test-label use, and output-level ensembling (including self-consistency). Only deterministic, non-learned post-processing (formatting, header insertion) may edit the final note.

BeTraC has two tracks that differ in architecture and total forward-pass parameters. The \textbf{Lightweight (LW)} track requires a single end-to-end model invoked exactly once per file, with no separately executed preprocessing or tools. Learned audio components (encoders, diarization, VAD) are allowed only as integrated submodules and count toward the 6B total parameter budget.
The \textbf{Heavyweight (HW)} track permits multi-model pipelines, agents, tools, and retrieval (medical ontologies, drug databases, guidelines, PubMed) under a 36B-parameter budget that sums all learned components in the forward pass. Audio tools (diarization, speaker embedding, VAD) may run as separate steps but must return only non-linguistic outputs (timestamps, speaker IDs, acoustic features).

\subsection{\textbf{Baselines}}
We provided per-track baselines that generate SOAP notes directly from audio without a separate ASR stage. The baselines are zero-shot: the models are prompted to produce a SOAP note with no fine-tuning or prompt optimization.
The \textbf{Lightweight} baseline is Qwen2.5-Omni-3B~\cite{Qwen2.5-Omni}; the \textbf{Heavyweight} baseline pool comprised Qwen2.5-Omni-3B/7B and Qwen3-Omni-30B-A3B (Instruct and Thinking)~\cite{xu2025qwen3}.

For reference, we also report ineligible cascaded systems that transcribe before summarizing: Whisper-large-v3 with Qwen3-30B-A3B reaches a development-set C-F1 of 0.288~\cite{yang2025qwen3technicalreport, radford2022robustspeechrecognitionlargescale}, and Qwen3-ASR-1.7B with Qwen3-30B-A3B reaches 0.281~\cite{shi2026qwen3asrtechnicalreport, yang2025qwen3technicalreport}. Because they pass an audio-derived transcript between components, these systems are ineligible for the BeTraC ranking.

\begin{table*}[t]
\centering
\caption{
ID and OOD evaluation results, ordered by ID C-F1.
Boldface indicates the best primary submission within each track and evaluation set for each metric.
$\dagger$ indicates a score that exceeds the corresponding reference-note score.
Gray $\pm$ values indicate 95\% confidence interval half-widths.
Hallucination is reported as a per-claim percentage, with lower values being better.
}
\label{tab:id_ood_results_with_judge}
\footnotesize
\setlength{\tabcolsep}{1.9pt}
\renewcommand{\arraystretch}{1.15}

\resizebox{\textwidth}{!}{%
\begin{tabular}{@{}l*{24}{c}@{}}
\toprule
\textbf{System}
& \multicolumn{14}{c}{\textbf{Automated metrics}}
& \multicolumn{10}{c}{\textbf{LLM-as-a-Judge}} \\
\cmidrule(lr){2-15}
\cmidrule(lr){16-25}
& \multicolumn{2}{c}{\textbf{C-F1}}
& \multicolumn{2}{c}{\textbf{C-P}}
& \multicolumn{2}{c}{\textbf{C-R}}
& \multicolumn{2}{c}{\textbf{R-2}}
& \multicolumn{2}{c}{\textbf{R-3}}
& \multicolumn{2}{c}{\textbf{R-L}}
& \multicolumn{2}{c}{\textbf{Words}}
& \multicolumn{2}{c}{\textbf{Faith.}}
& \multicolumn{2}{c}{\textbf{Cov.}}
& \multicolumn{2}{c}{\textbf{Struct.}}
& \multicolumn{2}{c}{\textbf{Conc.}}
& \multicolumn{2}{c}{\textbf{Halluc. (\%)}} \\
\cmidrule(lr){2-3}
\cmidrule(lr){4-5}
\cmidrule(lr){6-7}
\cmidrule(lr){8-9}
\cmidrule(lr){10-11}
\cmidrule(lr){12-13}
\cmidrule(lr){14-15}
\cmidrule(lr){16-17}
\cmidrule(lr){18-19}
\cmidrule(lr){20-21}
\cmidrule(lr){22-23}
\cmidrule(lr){24-25}
& \textbf{ID}
& \textbf{OOD}
& \textbf{ID}
& \textbf{OOD}
& \textbf{ID}
& \textbf{OOD}
& \textbf{ID}
& \textbf{OOD}
& \textbf{ID}
& \textbf{OOD}
& \textbf{ID}
& \textbf{OOD}
& \textbf{ID}
& \textbf{OOD}
& \textbf{ID}
& \textbf{OOD}
& \textbf{ID}
& \textbf{OOD}
& \textbf{ID}
& \textbf{OOD}
& \textbf{ID}
& \textbf{OOD}
& \textbf{ID}
& \textbf{OOD} \\
\midrule

\textit{Reference notes}
& --
& --
& --
& --
& --
& --
& --
& --
& --
& --
& --
& --
& 355.1
& 287
& 4.85\,{\color{gray}\tiny $\pm 0.03$}
& 4.85\,{\color{gray}\tiny $\pm 0.05$}
& 4.66\,{\color{gray}\tiny $\pm 0.05$}
& 4.27\,{\color{gray}\tiny $\pm 0.10$}
& 4.89
& 4.78
& 4.39
& 4.64
& 0.20
& 0.17 \\

\midrule
\multicolumn{25}{l}{\textbf{Baseline}} \\
\midrule

Qwen2.5-Omni-3B
& 0.309\,{\color{gray}\tiny $\pm 0.005$}
& 0.251\,{\color{gray}\tiny $\pm 0.008$}
& 0.343
& 0.261
& 0.295
& 0.262
& 0.116
& 0.079
& 0.050
& 0.029
& 0.189
& 0.144
& 817.4
& 935.2
& 2.92\,{\color{gray}\tiny $\pm 0.07$}
& 1.96\,{\color{gray}\tiny $\pm 0.08$}
& 3.63\,{\color{gray}\tiny $\pm 0.08$}
& 3.03\,{\color{gray}\tiny $\pm 0.11$}
& 2.95
& 2.67
& 2.28
& 2.27
& 8.91
& 19.72 \\

Qwen2.5-Omni-7B
& 0.326\,{\color{gray}\tiny $\pm 0.006$}
& 0.304\,{\color{gray}\tiny $\pm 0.009$}
& 0.379
& 0.324
& 0.296
& 0.301
& 0.140
& 0.103
& 0.061
& 0.039
& 0.224
& 0.180
& 448.2
& 722.1
& 3.75\,{\color{gray}\tiny $\pm 0.08$}
& 2.75\,{\color{gray}\tiny $\pm 0.11$}
& 4.04\,{\color{gray}\tiny $\pm 0.07$}
& 3.70\,{\color{gray}\tiny $\pm 0.10$}
& 2.97
& 2.77
& 3.00
& 2.60
& 5.13
& 14.52 \\

Qwen3-Omni-30B-Instruct
& 0.318\,{\color{gray}\tiny $\pm 0.004$}
& 0.293\,{\color{gray}\tiny $\pm 0.006$}
& 0.267
& 0.241
& 0.403
& 0.392
& 0.146
& 0.120
& 0.064
& 0.048
& 0.231
& 0.203
& 563.1
& 547.1
& 2.69\,{\color{gray}\tiny $\pm 0.05$}
& 2.42\,{\color{gray}\tiny $\pm 0.07$}
& 4.80$^\dagger$\,{\color{gray}\tiny $\pm 0.03$}
& 4.69$^\dagger$\,{\color{gray}\tiny $\pm 0.06$}
& 4.87
& 4.51
& 4.09
& 3.88
& 19.02
& 21.84 \\

Qwen3-Omni-30B-Thinking
& 0.285\,{\color{gray}\tiny $\pm 0.004$}
& 0.265\,{\color{gray}\tiny $\pm 0.006$}
& 0.239
& 0.214
& 0.363
& 0.361
& 0.133
& 0.116
& 0.055
& 0.044
& 0.219
& 0.196
& 509.8
& 479.7
& 2.54\,{\color{gray}\tiny $\pm 0.05$}
& 2.03\,{\color{gray}\tiny $\pm 0.06$}
& 4.76$^\dagger$\,{\color{gray}\tiny $\pm 0.03$}
& 4.73$^\dagger$\,{\color{gray}\tiny $\pm 0.05$}
& 4.86
& 4.44
& 4.03
& 3.89
& 17.78
& 24.39 \\

\midrule
\multicolumn{25}{l}{\textbf{Lightweight Track}} \\
\midrule

TalTech
& \textbf{0.543}\,{\color{gray}\tiny $\pm 0.006$}
& 0.505\,{\color{gray}\tiny $\pm 0.010$}
& 0.579
& \textbf{0.562}
& \textbf{0.518}
& 0.471
& 0.366
& 0.317
& 0.250
& 0.194
& 0.434
& 0.401
& 323.4
& 240.4
& \textbf{4.71}\,{\color{gray}\tiny $\pm 0.05$}
& \textbf{4.31}\,{\color{gray}\tiny $\pm 0.11$}
& \textbf{4.49}\,{\color{gray}\tiny $\pm 0.06$}
& 3.85\,{\color{gray}\tiny $\pm 0.10$}
& \textbf{4.88}
& \textbf{4.86}$^\dagger$
& 4.16
& 4.55
& 0.25
& 1.25 \\

NTT-HI-CS
& 0.540\,{\color{gray}\tiny $\pm 0.006$}
& \textbf{0.510}\,{\color{gray}\tiny $\pm 0.008$}
& \textbf{0.587}
& 0.555
& 0.507
& \textbf{0.479}
& \textbf{0.381}
& \textbf{0.348}
& \textbf{0.265}
& \textbf{0.219}
& \textbf{0.452}
& \textbf{0.435}
& 310.1
& 247.2
& 4.58\,{\color{gray}\tiny $\pm 0.06$}
& 3.91\,{\color{gray}\tiny $\pm 0.12$}
& 4.37\,{\color{gray}\tiny $\pm 0.06$}
& 3.64\,{\color{gray}\tiny $\pm 0.08$}
& 4.80
& 4.73
& 4.37
& 4.41
& 0.24
& 1.20 \\

KUSLP
& 0.515\,{\color{gray}\tiny $\pm 0.006$}
& 0.465\,{\color{gray}\tiny $\pm 0.008$}
& 0.557
& 0.509
& 0.488
& 0.437
& 0.367
& 0.330
& 0.253
& 0.206
& 0.433
& 0.422
& 318.9
& 248.8
& 4.59\,{\color{gray}\tiny $\pm 0.06$}
& 3.66\,{\color{gray}\tiny $\pm 0.13$}
& 4.46\,{\color{gray}\tiny $\pm 0.06$}
& 3.80\,{\color{gray}\tiny $\pm 0.09$}
& 4.72
& 4.73
& 4.11
& 4.44
& 0.38
& 3.89 \\

JYVA
& 0.501\,{\color{gray}\tiny $\pm 0.006$}
& 0.455\,{\color{gray}\tiny $\pm 0.008$}
& 0.543
& 0.458
& 0.473
& 0.461
& 0.351
& 0.312
& 0.243
& 0.195
& 0.412
& 0.376
& 323.0
& 303.7
& 4.61\,{\color{gray}\tiny $\pm 0.06$}
& 3.73\,{\color{gray}\tiny $\pm 0.12$}
& 4.25\,{\color{gray}\tiny $\pm 0.07$}
& \textbf{4.04}\,{\color{gray}\tiny $\pm 0.09$}
& 4.16
& 3.25
& 3.29
& 3.13
& \textbf{0.22}
& 2.09 \\

KIT-ISL-AI4LT
& 0.495\,{\color{gray}\tiny $\pm 0.006$}
& 0.462\,{\color{gray}\tiny $\pm 0.008$}
& 0.546
& 0.524
& 0.461
& 0.423
& 0.360
& 0.319
& 0.250
& 0.201
& 0.432
& 0.407
& 291.5
& 220.3
& 4.62\,{\color{gray}\tiny $\pm 0.05$}
& 4.06\,{\color{gray}\tiny $\pm 0.11$}
& 4.37\,{\color{gray}\tiny $\pm 0.06$}
& 3.95\,{\color{gray}\tiny $\pm 0.10$}
& 4.83
& 4.84$^\dagger$
& \textbf{4.56}$^\dagger$
& \textbf{4.77}$^\dagger$
& 0.27
& \textbf{0.73} \\

DirectSense
& 0.434\,{\color{gray}\tiny $\pm 0.006$}
& 0.448\,{\color{gray}\tiny $\pm 0.008$}
& 0.541
& 0.482
& 0.369
& 0.428
& 0.249
& 0.241
& 0.153
& 0.134
& 0.331
& 0.321
& 240.9
& 246.1
& 4.13\,{\color{gray}\tiny $\pm 0.07$}
& 3.53\,{\color{gray}\tiny $\pm 0.12$}
& 3.93\,{\color{gray}\tiny $\pm 0.07$}
& 3.96\,{\color{gray}\tiny $\pm 0.10$}
& 3.93
& 4.28
& 3.45
& 3.75
& 2.29
& 4.25 \\

ASLP
& 0.397\,{\color{gray}\tiny $\pm 0.006$}
& 0.382\,{\color{gray}\tiny $\pm 0.007$}
& 0.393
& 0.345
& 0.413
& 0.442
& 0.222
& 0.192
& 0.134
& 0.103
& 0.305
& 0.282
& 437.5
& 407.8
& 2.18\,{\color{gray}\tiny $\pm 0.08$}
& 1.19\,{\color{gray}\tiny $\pm 0.06$}
& 3.30\,{\color{gray}\tiny $\pm 0.08$}
& 3.03\,{\color{gray}\tiny $\pm 0.09$}
& 3.92
& 4.08
& 3.81
& 3.68
& 18.82
& 34.10 \\

IASP
& 0.362\,{\color{gray}\tiny $\pm 0.006$}
& 0.347\,{\color{gray}\tiny $\pm 0.007$}
& 0.406
& 0.381
& 0.335
& 0.329
& 0.130
& 0.113
& 0.080
& 0.066
& 0.181
& 0.158
& 743.0
& 728.1
& 2.19\,{\color{gray}\tiny $\pm 0.08$}
& 1.47\,{\color{gray}\tiny $\pm 0.06$}
& 2.70\,{\color{gray}\tiny $\pm 0.08$}
& 2.50\,{\color{gray}\tiny $\pm 0.07$}
& 2.64
& 3.08
& 1.19
& 1.19
& 18.75
& 35.31 \\

\midrule
\multicolumn{25}{l}{\textbf{Heavyweight Track}} \\
\midrule

TalTech
& \textbf{0.563}\,{\color{gray}\tiny $\pm 0.006$}
& \textbf{0.556}\,{\color{gray}\tiny $\pm 0.008$}
& \textbf{0.601}
& 0.571
& \textbf{0.536}
& \textbf{0.549}
& \textbf{0.401}
& \textbf{0.381}
& \textbf{0.280}
& \textbf{0.242}
& \textbf{0.468}
& \textbf{0.468}
& 312.2
& 257.1
& 4.80\,{\color{gray}\tiny $\pm 0.04$}
& \textbf{4.63}\,{\color{gray}\tiny $\pm 0.08$}
& 4.50\,{\color{gray}\tiny $\pm 0.06$}
& 4.14\,{\color{gray}\tiny $\pm 0.10$}
& \textbf{4.91}$^\dagger$
& 4.80$^\dagger$
& \textbf{4.41}$^\dagger$
& 4.59
& \textbf{0.08}$^\dagger$ 
& \textbf{0.30} \\

KUSLP
& 0.544\,{\color{gray}\tiny $\pm 0.006$}
& 0.530\,{\color{gray}\tiny $\pm 0.009$}
& 0.596
& \textbf{0.580}
& 0.509
& 0.494
& 0.393
& 0.374
& 0.277
& 0.241
& 0.459
& \textbf{0.468}
& 310.6
& 248.0
& \textbf{4.81}\,{\color{gray}\tiny $\pm 0.04$}
& 4.53\,{\color{gray}\tiny $\pm 0.09$}
& 4.32\,{\color{gray}\tiny $\pm 0.06$}
& 3.67\,{\color{gray}\tiny $\pm 0.09$}
& 4.82
& 4.84$^\dagger$
& 4.34
& \textbf{4.76}$^\dagger$
& 0.17$^\dagger$
& 0.42 \\

NTT-HI-CS
& 0.539\,{\color{gray}\tiny $\pm 0.006$}
& 0.517\,{\color{gray}\tiny $\pm 0.009$}
& 0.579
& 0.561
& 0.511
& 0.485
& 0.377
& 0.348
& 0.261
& 0.218
& 0.448
& 0.440
& 315.1
& 246.9
& 4.47\,{\color{gray}\tiny $\pm 0.06$}
& 3.60\,{\color{gray}\tiny $\pm 0.12$}
& 4.38\,{\color{gray}\tiny $\pm 0.06$}
& 3.65\,{\color{gray}\tiny $\pm 0.10$}
& 4.73
& 4.68
& 4.30
& 4.34
& 0.41
& 2.06 \\

ASLP
& 0.506\,{\color{gray}\tiny $\pm 0.006$}
& 0.501\,{\color{gray}\tiny $\pm 0.008$}
& 0.504
& 0.481
& 0.518
& 0.532
& 0.341
& 0.295
& 0.227
& 0.168
& 0.415
& 0.405
& 365.6
& 333.7
& 4.36\,{\color{gray}\tiny $\pm 0.07$}
& 4.41\,{\color{gray}\tiny $\pm 0.10$}
& \textbf{4.69}$^\dagger$\,{\color{gray}\tiny $\pm 0.05$}
& 4.38$^\dagger$\,{\color{gray}\tiny $\pm 0.10$}
& 4.79
& \textbf{4.85}$^\dagger$
& 4.31
& 3.96
& 0.94
& 0.79 \\

Slate Lab
& 0.475\,{\color{gray}\tiny $\pm 0.010$}
& 0.462\,{\color{gray}\tiny $\pm 0.011$}
& 0.510
& 0.465
& 0.467
& 0.475
& 0.300
& 0.294
& 0.206
& 0.182
& 0.357
& 0.370
& 471.5
& 365.2
& 4.51\,{\color{gray}\tiny $\pm 0.08$}
& 4.26\,{\color{gray}\tiny $\pm 0.11$}
& 4.08\,{\color{gray}\tiny $\pm 0.12$}
& \textbf{4.45}$^\dagger$\,{\color{gray}\tiny $\pm 0.13$}
& 3.91
& 4.16
& 3.07
& 3.89
& 2.15
& 1.77 \\

\bottomrule
\end{tabular}
}
\end{table*}

\subsection{\textbf{Competition Evaluation}}
The official evaluation compares the SOAP notes generated by each system with the reference SOAP notes (produced by the Synth-DoPaCo Kimi K2 pipeline). Teams were provided with reference SOAP notes for the training set and code to compute these metrics.

\subsubsection{\textbf{Open Medical Concept F1}}

Open Medical Concept F1 (C-F1) is the primary ranking metric. The default matcher combines keyword matching over a curated MeSH vocabulary~\cite{lipscomb2000medical} with biomedical named-entity recognition, and the resulting concept identifiers are compared as sets between the generated and reference notes. For dialogue $i$, let $C_i^{\mathrm{pred}}$ and $C_i^{\mathrm{ref}}$ be the concept-identifier sets of the generated and reference notes and $O_i=\lvert C_i^{\mathrm{pred}}\cap C_i^{\mathrm{ref}}\rvert$ their overlap. Precision and recall are

\noindent
\begin{minipage}[t]{0.48\linewidth}
\begin{equation}
P_i =
\frac{\left|C_i^{\mathrm{pred}} \cap C_i^{\mathrm{ref}}\right|}
     {\left|C_i^{\mathrm{pred}}\right| + \epsilon},
\end{equation}
\end{minipage}%
\hfill
\begin{minipage}[t]{0.48\linewidth}
\begin{equation}
R_i =
\frac{\left|C_i^{\mathrm{pred}} \cap C_i^{\mathrm{ref}}\right|}
     {\left|C_i^{\mathrm{ref}}\right| + \epsilon},
\end{equation}
\end{minipage}

\vspace{6pt}

\noindent where \(\epsilon = 10^{-6}\). 

The C-F1 score for dialogue \(i\) and the final macro-averaged score over
all \(N\) evaluation dialogues are computed as
\begin{equation}
F_i = \frac{2P_iR_i}{P_i + R_i + \epsilon} 
\qquad \text{and} \qquad
\mathrm{Concept\ F1} = \frac{1}{N}\sum_{i=1}^{N} F_i.
\end{equation}

Concept precision and concept recall are also reported as macro averages over all evaluation dialogues. Because the extracted concept identifiers are represented as sets, repeated occurrences of the same concept within a note are counted only once.

\subsubsection{\textbf{ROUGE}}
ROUGE measures textual overlap between each generated SOAP note and its reference note \cite{lin-2004-rouge, rouge-score}. We report ROUGE 2 F1, ROUGE 3 F1, and ROUGE L F1 as secondary metrics. ROUGE 2 and ROUGE 3 measure bigram and trigram
overlap, respectively. ROUGE L is based on the longest common subsequence. Each score is first computed for each dialogue and then macro averaged over the evaluation set. 

\subsection{\textbf{Post-competition Evaluation}}
\label{sec:ood}
We conducted additional post competition analysis
with out-of-domain human-acted doctor patient conversations and by using LLM-as-a-judge SOAP note evaluation (Table~\ref{tab:dataset_statistics}). These results did not affect the outcome of the competition.

The out-of-distribution (OOD) conversations came from two sources. First, we used 272 human-roleplayed doctor–patient conversation audios in the format of Objective Structured Clinical Examinations (OSCEs) focusing on respiratory cases~\cite{Fareez2022}, which provide ground-truth transcripts and are designed for high audio quality and few disfluencies. Second, we used three real-world human-acted conversations recorded at the JSALT 2025 workshop, including a speaker with a substantial accent and captured with a single microphone. For both, transcripts were produced semi-automatically, and reference SOAP notes were generated with the same Kimi K2 Thinking pipeline as Synth-DoPaCo. 
Neither source was provided as BeTraC training data; the JSALT recordings, moreover, are private and were never released, so they cannot have been seen by any LALM, giving a contamination-free OOD test.

For LLM-as-a-judge evaluation, we use the two-stage pipeline developed for Synth-DoPaCo, with Gemma 4 31B IT as the judge model~\cite{gemma4modelcard} (distinct from the Kimi K2 ground truth note-construction model, to avoid self-preference bias). 
We run inference in FP8 precision using vLLM~\cite{vllm}, with the temperature set to 0 and maximum generation lengths of 16,000 and 12,000 tokens for claim extraction and final judgment, respectively. The LLM-as-a-judge evaluation required approximately 23 H100 GPU-seconds per dialogue.
\footnote{ In experiments on notes produced by the baseline model, we found that the LLM gets slightly more generous as it is quantized (within 0.1) on each rating dimension. It is additionally substantially more generous (+0.4) when run with vLLM compared to Ollama, for unknown reasons.} This pipeline first performs claim extraction from the SOAP note. Then, in the second stage, the judge model evaluates the \textit{Faithfulness, Coverage, Structure}, and \textit{Conciseness} of the claim list relative to the dialogue transcript. Faithfulness evaluates claim support and contradictions. Coverage evaluates the completeness of documentation relative to transcript evidence. Structure evaluates SOAP formatting and section placement. Conciseness evaluates redundancy and low value content. We consider faithfulness and coverage to be the most important properties of SOAP notes, with structure and conciseness included for their roles in readability.

\begin{table*}[t]
\centering
\caption{Design space of the final BeTraC submissions. A split backbone swatch
denotes different LW\,/\,HW models. \cmark\ in Ext./Aug.\ marks use
of external clinical corpora; (TTS) marks synthetic-speech augmentation. In
Struct./Tools, (encoder) marks diarization integrated at the audio encoder and (tool)
an external diarization step. A dash denotes a technique not used. Backbone colors:
\swone{cqwen25}~Qwen2.5-Omni, \swone{cvoxtral}~Voxtral, \swone{cqwen3o}~Qwen3-Omni-30B.
\textsuperscript{\dag}~LW only; \textsuperscript{\ddag}~Slate Lab's contrast
system only.}
\label{tab:design_space}
\setlength{\tabcolsep}{5pt}
\renewcommand{\arraystretch}{1.1}
\resizebox{0.72\textwidth}{!}{%
\begin{tabular}{@{}lllllllll@{}}
\toprule
\textbf{System} & \textbf{Track} & \textbf{Backbone} & \textbf{Params}
& \textbf{Adapt.} & \textbf{RL/Pref} & \textbf{Ext./Aug.} & \textbf{Long-ctx} & \textbf{Struct./Tools}\\
\midrule
TalTech~\cite{TalTech}        & L,H & \swone{cvoxtral}~Voxtral 3B / 24B                       & 3/24B    & full     & GSPO & \dna               & \dna           & \dna \\
KUSLP~\cite{KUSLP1,KUSLP2}         & L,H & \swtwo{cqwen25}{cqwen3o}~Qwen2.5-Omni / Qwen3-Omni-30B & 3B / 30B & LoRA    & GRPO & \cmark\,{\scriptsize(TTS)} & time-compress & \dna \\
NTT-HI-CS~\cite{NTT}      & L,H & \swone{cqwen25}~Qwen2.5-Omni-3B                        & 4.0B     & full     & DAPO & \cmark\textsuperscript{\dag} & avg-pool & \dna \\
ASLP~\cite{ASLP}           & L,H & \swtwo{cqwen25}{cqwen3o}~Qwen2.5-Omni / Qwen3-Omni-30B & 4B / 30B & full  & GRPO & \cmark            & \dna           & \dna \\
JYVA~\cite{JYVA}           & L   & \swone{cqwen25}~Qwen2.5-Omni + DiariZen                & 6.0B     & QFormer & \dna  & \cmark            & \dna           & diar (encoder) \\
KIT-ISL-AI4LT~\cite{KIT}  & L   & \swone{cqwen25}~Qwen2.5-Omni-3B                        & 4.0B     & LoRA    & \dna  & \cmark            & \dna     & \dna \\
Slate Lab      & H   & \swone{cqwen3o}~Qwen3-Omni-30B                         & 35B      & LoRA    & \dna  & \dna               & \dna           & diar (tool)\textsuperscript{\ddag} \\
DirectSense    & L   & \swone{cvoxtral}~Voxtral-Mini-3B                       & 4.7B     & prompt-only & \dna & \dna           & \dna           & \dna \\
IASP           & L   & \swone{cqwen25}~Qwen2.5-Omni-3B                        & 4.1B     & QLoRA   & \dna  & \dna               & \dna           & \dna \\
\bottomrule
\end{tabular}
}
\end{table*}

\section{\textbf{Results}}
\subsection{\textbf{Submitted System Descriptions}}
Nine teams participated (eight in LW, five in HW, four in both) and could also submit off-leaderboard contrast systems. Table~\ref{tab:design_space} summarizes their designs; we highlight the trends here and analyze their effect in Section~\ref{sec:discussion}.

Every system adapted a pretrained open audio-language model rather than training one from scratch: Qwen2.5-Omni-3B~\cite{Qwen2.5-Omni} was the most common backbone (six teams), with Voxtral~\cite{liu2025voxtral} and (HW-only) Qwen3-Omni-30B-A3B~\cite{xu2025qwen3} as alternatives. Teams differed mainly in how they adapted the backbone rather than which one they chose. Most used full fine-tuning or LoRA/QLoRA~\cite{Hu2021LoRALA,Dettmers2023QLoRAEF}, while JYVA instead trained a QFormer~\cite{li2023blip} to align diarization-aware audio features. Four teams then added metric-aligned RLVR rewarding C-F1 (and sometimes ROUGE or SOAP-format terms): TalTech (GSPO~\cite{Zheng2025GroupSP}), NTT-HI-CS (DAPO~\cite{Yu2025DAPOAO}), and KUSLP and ASLP (GRPO~\cite{Shao2024DeepSeekMathPT}). DirectSense did no fine-tuning at all, prompt-tuning a frozen Voxtral-Mini-3B~\cite{liu2025voxtral}. All teams trained on Synth-DoPaCo, four supplementing it with external corpora~\cite{yim2023acibench,ben-abacha-etal-2023-empirical,papadopoulos-korfiatis-etal-2022-primock57} and KUSLP additionally augmenting the audio via TTS re-rendering. With conversations averaging nine minutes,
three teams added explicit long-context mechanisms in order to shorten the input sequence and thus reduce computational burden: feature average-pooling
(NTT-HI-CS) and waveform time-compression (KUSLP).

Notably, the Heavyweight track's headline affordances went almost unused: no system used retrieval, ontologies, or tool-calling agents---all teams generated the note in a single pass. The only exception was speaker diarization, used by two systems: JYVA, which integrates DiariZen~\cite{han2025leveraging} at the audio encoder, and Slate Lab's contrast system, which runs a separate Pyannote step. All post-processing was deterministic header/format cleanup, with no learned model editing the note.

\subsection{\textbf{Metric Scores}}
\begin{figure*}[t]
    \centering

    \begin{minipage}[t]{0.2\textwidth}
        \centering
        \includegraphics[width=\linewidth]
        {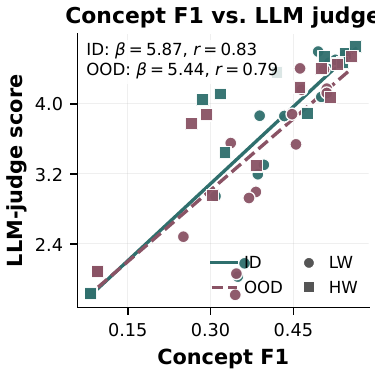}
        \vspace{-1mm}

        \textbf{(a)}
    \end{minipage}
    \hfill
    \begin{minipage}[t]{0.2\textwidth}
        \centering
        \includegraphics[width=\linewidth]
        {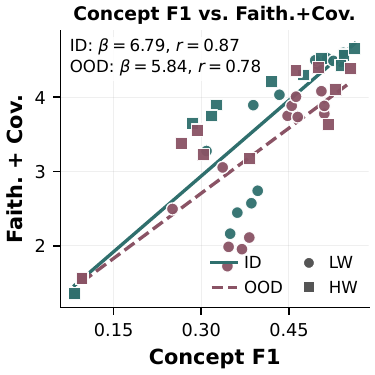}
        \vspace{-1mm}

        \textbf{(b)}
    \end{minipage}
    \hfill
    \begin{minipage}[t]{0.2\textwidth}
        \centering
        \includegraphics[width=\linewidth]
        {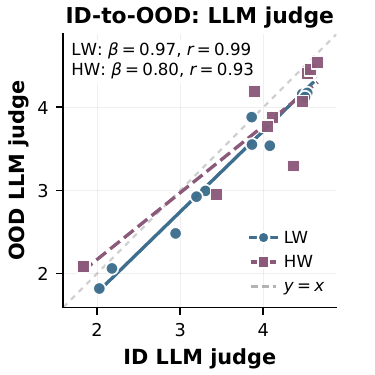}
        \vspace{-1mm}

        \textbf{(c)}
    \end{minipage}
    \hfill
    \begin{minipage}[t]{0.2\textwidth}
        \centering
        \includegraphics[width=\linewidth]
        {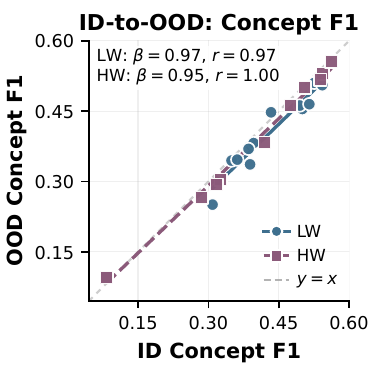}
        \vspace{-1mm}

        \textbf{(d)}
    \end{minipage}
    
    \caption{
    Relationships between evaluation performance on the ID and OOD subsets across primary submissions, contrast systems, and baselines.
    (a) C-F1 vs. the aggregate LLM-as-a-judge score on ID and
    OOD data.
    (b) C-F1 vs. the average of faithfulness and coverage on
    ID and OOD data.
    (c) ID vs. OOD LLM-as-a-judge performance.
    (d) ID vs. OOD C-F1 performance.
    Marker shape distinguishes LW and HW systems.
    Regression slopes and Pearson correlations are shown within each panel.
    }
    \label{fig:metric_relationships}
    \vspace{-0.2cm}
\end{figure*}

Table~\ref{tab:id_ood_results_with_judge} reports the results for each team, baseline, and reference on each metric and Figure~\ref{fig:metric_relationships} shows the relationships across metrics and datasets.
The reference notes are judged as the strongest by the LLM judge. They are particularly strong in faithfulness (4.85 ID, OOD) and structure (4.89 ID, 4.78 OOD). They are judged to be weaker in coverage (4.66 ID, 4.27 OOD) and conciseness (4.39 ID, 4.64 OOD). 

TalTech won both the LW and HW tracks of the challenge. Their systems were ranked first in C-F1 in all settings except LW OOD, where they ranked second. Their LW system would be the likely LW pick of the LLM judge, receiving the highest faithfulness scores both ID and OOD and having generally high other scores. Their HW system was possibly the strongest in the competition overall. It is highly faithful (4.80 ID, 4.63 OOD), had the lowest hallucination rates (0.08\% ID, 0.30\% OOD), and scored well in structure and conciseness, often exceeding the reference notes. However, it suffers from weaker coverage, especially OOD (4.14), which we will discuss in more detail.

NTT-HI-CS received second place in LW and third place in HW. The LW track was extremely close---NTT-HI-CS's LW system had C-F1 of just 0.003 lower than TalTech, and, in LW OOD, they had higher C-F1 than TalTech. Their ROUGE and Concept Recall scores were the highest in LW, both ID and OOD. Their LLM judge scores were weaker than TalTech in all settings except conciseness in LW ID.

KUSLP received third place in LW and second in HW. Their HW system was highly faithful, receiving the highest score ID and second highest OOD, and produced well structured and concise notes, but suffered from weak coverage, especially OOD (3.67).

The HW system of ASLP deserves special mention. Despite having lower Concept and ROUGE scores ID and OOD, it was highly regarded by the LLM judge, where it received the highest coverage score ID and second highest OOD, and the highest structure score OOD. This system produces longer notes (15--30\% longer than TalTech's) that the LLM judged as more comprehensive but less faithful. We discuss this case further in the next section.

\section{\textbf{Discussion}}
\label{sec:discussion}

TalTech, NTT-HI-CS, KUSLP, and ASLP (HW) built the strongest systems, and all four used the same recipe: full or LoRA fine-tuning, then some form of RLVR, on a Qwen or Voxtral base model. This was the winning approach in the challenge. What set the top teams apart was mostly how they adapted the model, not which base model they picked: the two best Lightweight systems use different base models yet score within 0.003 C-F1 of each other.

In Fig.~\ref{fig:metric_relationships}, we examine the relationships between C-F1 metric vs. the average LLM as a judge score, as well as metric scores transfer from ID to OOD, using all systems including contrasts. We find a clear pattern. C-F1 transfers near perfectly from ID to OOD~(Fig.~\ref{fig:metric_relationships}d), with $r \ge 0.97$. LLM judge scores transfer a bit less well, with $r \ge 0.93$~(Fig.~\ref{fig:metric_relationships}c). In each case, we see no clear signs of overfitting to ID, with outlying points on both sides of the line (e.g., Slate Lab received an average LLM judge score of 3.9 ID and 4.2 OOD).
These findings are notable given that the ID audio are entirely synthetic, and OOD are real recorded conversations. This suggests that synthetic data generation, and particularly RLVR in synthetic environments, is a viable path forward to improving the general long context summarization capabilities of LALMs. This near-linear ID $\rightarrow$ OOD relationship echoes the "accuracy on the line" phenomenon, where OOD performance is strongly linearly predicted by ID performance across a wide range of models~\cite{miller2021}. Our results extend the observation to generative clinical summarization and, notably, to a synthetic-to-real distribution shift.

The clearest signs of weakness in transferability are between C-F1 and the LLM judge scores (Fig.~\ref{fig:metric_relationships}a-b), which we find to be attributable to at least four factors: C-F1 does not give credit for readability, it does not penalize repetition, it handles negated concepts incorrectly, and errors in the reference note propagate to C-F1 but not the LLM judge. We take these in turn.
In Fig.~\ref{fig:metric_relationships}a, we see two clusters of points with LLM judge score of around 2 that are separated by $\sim 0.2$ points of C-F1. The group on the left, with low C-F1, are notes that have no relationship to the dialogue but still receive high LLM judge readability scores. The group on the right are highly repetitive notes, which are judged poorly for conciseness but still receive high C-F1 (which is invariant to repetition).
Removing the structure and conciseness judge dimensions tightens the fit on ID data, but not on OOD data (Fig.~\ref{fig:metric_relationships}b).

A salient disagreement between the automated metrics and the judge is HW TalTech vs.\ ASLP. ASLP scores higher on judged coverage, both ID and OOD, yet lower on every automated metric, including Concept Recall, which is supposed to reward capturing more of the reference's concepts. The main cause is negation. ASLP documents extensive pertinent negatives (``Denies cough, chest pain, \dots''), but the reference notes contain few negatives. In addition, the C-F1 matcher does not handle negation: it extracts ``cough'' and ``chest pain'' as \emph{present} concepts. The result is that including negatives tends to reduce Concept Precision and F1, even if they are supported by the transcript.
Negations account for roughly 57\% of ASLP's precision gap and 35\% of its F1 gap against TalTech. The LLM judge does not see the reference note and rewards supported negative statements with higher coverage scores.
Thus, ASLP HW is a stronger system than C-F1 implies, due to peculiarities of the automated metrics.

\section{\textbf{Conclusion}}

These results suggest that end-to-end LALMs can produce faithful SOAP notes from long clinical audio, and, more broadly, that training and reward-optimizing on synthetic audio can transfer to real recordings---offering a scalable route to long-context audio understanding that sidesteps the data-scarcity and privacy barriers of real clinical speech.

Several caveats temper this. The reference notes are a strong comparator but not a perfect ceiling. The best systems do not quite match their faithfulness (the one dimension on which the references remain unbeaten), yet several systems equal or exceed them on coverage, structure, or conciseness. The references are themselves imperfect and could be improved; for instance, they appear to under-document pertinent negatives relative to some submissions (\S~\ref{sec:discussion}), so exceeding them on a given dimension does not certify a note as clinically ideal. More fundamentally, our synthetic audio does not yet reproduce the non-lexical cues (e.g., prosody, hesitation, audible symptoms) that motivate end-to-end modeling in the first place. These are challenging to simulate with current open-weight technology. On the metric side, no single automated score is fully trustworthy: C-F1 is cheap and reproducible but blind to negation, repetition, and readability, while the LLM judge captures these but brings its own biases and cost (and, as we observe, sensitivity to quantization and serving stack).

Finally, the Heavyweight track's headline affordances (retrieval, tools, and agents) went almost entirely unused. We read this less as a lack of interest than as a sign that audio-space processing tools remain far less mature than their text counterparts, and are hard to employ without crossing the no-transcription rule.

\section*{Acknowledgment}
This work was supported by the 2025 Jelinek Memorial Summer Workshop on Speech and Language Technologies (JSALT 2025), hosted at the Brno University of Technology and organized by the Center for Language and Speech Processing (CLSP) at Johns Hopkins University. We acknowledge the resources of the Ohio Supercomputer Center~\cite{OhioSupercomputerCenter1987}.

\textbf{AI-Generated Content Disclosure:} AI tools were used to assist with language editing and grammar enhancement across the manuscript. The authors reviewed the final text and take full responsibility for it.

\bibliographystyle{IEEEtran}
\bibliography{references}

\end{document}